\documentclass[a4paper]{article}

\usepackage{INTERSPEECH2022}
\usepackage{multirow}

\title{The DKU-Tencent System for the VoxCeleb Speaker Recognition Challenge 2022}
\name{Xiaoyi Qin$^{1,2*}$ Na Li$^{2*}$, Yuke Lin$^{1*}$\thanks{* Equal contribution. The work was done at Tencent during the internship for the first author}, Yiwei Ding$^{2}$, Chao Weng$^{2}$, Dan Su$^{2}$, Ming Li$^{1}$}
\address{$^{1}$ Data Science Research Center, Duke Kunshan University, Kunshan, China\\
		$^{2}$ Tencent AI Lab, Shenzhen, China}
		
\email{ming.li@whu.edu.cn, 2020102110042@whu.edu.cn}

\begin{document}

\maketitle
\begin{abstract}

This paper is the system description of the DKU-Tencent System for the VoxCeleb Speaker Recognition Challenge 2022 (VoxSRC22). In this challenge, we focus on track1 and track3. For track1, multiple backbone networks are adopted to extract frame-level features. Since track1 focus on the cross-age scenarios, we adopt the cross-age trials and perform QMF to calibrate score. The magnitude-based quality measures achieve a large improvement. For track3, the semi-supervised domain adaptation task, the pseudo label method is adopted to make domain adaptation. Considering the noise labels in clustering, the ArcFace is replaced by Sub-center ArcFace. The final submission achieves 0.107 mDCF in task1 and 7.135\% EER in task3.
\end{abstract}


\section{Task1: Fully supervised speaker verification}

\subsection{Data Augmentation}
We adopt the on-the-fly data augmentation \cite{on-the-fly} to add additive background noise or convolutional reverberation noise for the time-domain waveform. The MUSAN \cite{musan} and RIR Noise \cite{RIR} datasets are used as noise sources and room impulse response functions, respectively. The convolution operation is performed for the reverberation with 40,000 simulated room impulse responses. We only use RIRs from small and medium rooms. To further diversify training samples, we apply amplification or playback speed change (pitch remains untouched) to audio signals. Also, we apply speaker augmentation with speed perturbation \cite{dku_voxsrc20}. We speed up or down each utterance by a factor of 0.9 or 1.1, yielding shifted pitch utterances that are considered from new speakers. As a result, the training data includes $3,276,027 (1,092,009\times3)$ utterances from $17,982 (5,994\times3)$ speakers. The probability of Noise/RIR/Tempo/Volume is 0.66. 

\subsection{Speaker Embedding Model}

Generally, the speaker embedding model consists of three parts: backbone network, encoding layer and loss function. The ArcFace \cite{arcface}, which could increase intra-speaker distances while ensuring inter-speaker compactness, is used as a loss function. The backbone and encoding layer are reported as follows.

\subsubsection{Backbone Network}
In this module, we introduce four different speaker verification systems, including the SimAM-ResNet34\cite{simam}, the ResNet101\cite{resnet}, ResNet152, and Res2Net101\cite{res2net}. The acoustic features are 80-dimensional log Mel-filterbank energies with a frame length of 25ms and hop size of 10ms. The extracted features are mean-normalized before feeding into the deep speaker network.

\subsubsection{Encoding Layer}

The encoding layer could transform the variable length feature maps into a fixed dimension vector, that is, the frame-level feature is converted to the segment-level feature. In this module, we adopt the statistic pooling\cite{xvector}, standard deviation pooling and attentive statistic pooling\cite{asp_pooling}. The statistic pooling layer computes the mean and standard deviation of the output feature maps along with the frequency dimension. The standard deviation pooling layer only calculates the standard deviation. 

\subsection{Training Strategy and Implement Detail}

The speaker embedding model is trained through two stages. In the first stage, the baseline system adopts the SGD optimizer. We adopt the ReduceLROnPlateau scheduler with 2 epochs of patience. The init learning rate starts from 0.1, the minimum learning rate is 1.0e-4, and the decay factor is 0.1. The margin and scale of ArcFace are set as 0.2 and 32, respectively. We perform a linear warmup learning rate schedule at the first 2 epochs to prevent model vibration and speed model training. The input frame length is fixed at 200 frames. In the second stage, we following the SpeakIn training protocol\cite{speakin21} with large margin fine-tune (LMFT)\cite{idlab_voxsrc20}. In the LMFT stage, we also remove the speaker augmentation and reduce the probability of data augmentation to 0.33. What's more, the margin is increased from 0.2 to 0.5. According to the speaker embedding model size and the memory limit of GPU, the frame length is expended from 200 to 400 or 600. The learning rate decays from 1.0e-4 to 1.0e-5.

\begin{table*}[htbp]\centering \scriptsize 
    \caption{The baseline system performance of track1.}
     \label{tab:baseline_system}
    \begin{tabular}{lccccccccc}
    \toprule
	\multirow{2}*{\textbf{Model}} & \multicolumn{2}{c}{\textbf{Vox-O(cleaned)}} & \multicolumn{2}{c}{\textbf{VoxSRC22 val}} & \multicolumn{2}{c}{\textbf{Vox-CA}} & \multicolumn{2}{c}{\textbf{VoxSRC22 eval}} \\
	\cmidrule(lr){2-3} \cmidrule(lr){4-5} \cmidrule(lr){6-7} \cmidrule(lr){8-9} & \textbf{EER[\%]} & \textbf{mDCF$_{0.05}$} & \textbf{EER[\%]} & \textbf{mDCF$_{0.05}$} & \textbf{EER[\%]} & \textbf{mDCF$_{0.05}$} & \textbf{EER[\%]} & \textbf{mDCF$_{0.05}$} \\
    \midrule 
    SimAM-ResNet34 \\
    + LMFT & 0.542 & 0.034 & 1.789 & 0.108 & 6.944 & 0.263 & 3.585 & 0.219 \\
    ++ AS-Norm (Top 400) & - & - & - &  - & 6.852 & 0.249 & 3.400 & 0.200\\
    +++ QMF & - & - & - & - & 6.801 & 0.240 & 2.478 & 0.139 \\
    \bottomrule
    
    \end{tabular}
    \end{table*}

\subsection{Score Calibration and Normalization}

The cosine similarity is the back-end scoring method. After scoring, results from all trials are subject to score normalization. We utilize Adaptive Symmetric Score Normalization (AS-Norm) \cite{asnorm} in our systems. The imposter cohort consists of the average of the length normalized utterance-based embeddings of each training speaker, i.e., a speaker-wise cohort with 5994 embeddings.

Quality Measure Functions could calibrate the scores and improve the system performance. As \cite{idlab_voxsrc21} described, we adopt the two qualities set: speech duration and magnitude rate. For speech duration, the value of $|log(d_{u}-d_{min})|$ is adopt as a QMF. $d_{u}$ indicates the duration of enrollment or test utterance. $d_{min}$ is slightly lower than the minimum test utterance duration to prevent the $log(.)$ approaching NaN. The magnitude rate is formula as:$|log(\frac{\Vert \mathbf{z}_{e}\Vert}{\Vert \mathbf{z}_{t}\Vert})|$. As shown in \cite{ringloss}, the embedding magnitude could contain quality information about the original utterance. The experimental result also verifies the magnitude information is effective.

\subsection{Results}

\subsubsection{Baseline System}
In this subsection, we will show baseline system performance and how to improve performance step by step. We adopt the SimAM-ResNet34 as our baseline system, which is easy to train and fast to fit. At first, we validate the baseline system under the VoxSRC22 validation and evaluation set. It's easily observed that there is a large gap between the val set and eval set. To prevent the model overfit the validation set, we follow our previous work \cite{casv}, which is focused on the cross-age speaker verification, and adopt the Vox-CA15 trial as our validate trial. The design detail is shown in the \cite{casv}. Therefore, tuning the hyper-parameter is operated on the Vox-CA. The AS-Norm achieves a relative 10\% improvement, but the QMF achieves a 30\% improvement. According to the weight of logistic regression, the coefficient of magnitude rate is much higher than that of the similarity score between enrollment and test embedding. After all back-end processing, the system achieves the relative 35\% improvement in mDCF. The other models will adopt the same operation, and Table.\ref{tab:final_system_task1} only reports the final results.  

\subsubsection{Fusion System}   
Table.\ref{tab:final_system_task1} presents the performance of final models we used. At first pre-training stage, we have multiple combinations of the backbone network and encoding layer, the models in Table.\ref{tab:final_system_task1} are the best ones on VoxCeleb-O (cleaned) trials. Comparing models 1 and 2, we find the lower EER in Vox-CA and the better mDCF in VoxSRC22 eval. Therefore, the final score is the sum of models 2, 3, 4 and 5  with average weight. The final result achieves the 2.078\% EER and 0.107 mDCF.  
    
\begin{table}[htbp]\centering \scriptsize 
    \caption{The performance of various systems in the track3.}
     \label{tab:final_system_task1}
    \begin{tabular}{lccccccc}
    \toprule
    \multirow{2}*{\textbf{ID \& Model}} &  \multicolumn{2}{c}{\textbf{Vox-CA}} & \multicolumn{2}{c}{\textbf{VoxSRC22 eval}} \\
    \cmidrule(lr){2-3} \cmidrule(lr){4-5} \cmidrule(lr){6-7} & \textbf{EER[\%]} & \textbf{mDCF$_{0.05}$} & \textbf{EER[\%]} & \textbf{mDCF$_{0.05}$}  \\
    \midrule 

    1 SimAM-ResNet34 & 6.801 & 0.240 & 2.478 & 0.139\\
    2 ResNet101-ASP  & 6.494 & 0.224 & 2.139 & 0.122\\
    3 ResNet152-Stat & 6.654 & 0.240 & - & - \\
    4 ResNet152-ASP  & 6.538 & 0.224 & - & - \\
    5 Res2Net101-Std & 6.534 & 0.244 & - & - \\
    
    \midrule
    Fusion(1+2+3+4) & - & - & 2.085 & 0.117\\
    Fusion(2+3+4+5) & - & - & 2.078 & 0.107\\
    
    \bottomrule
     
    \end{tabular}
    \end{table}

\begin{figure}
  \centering
  \includegraphics[width=\linewidth]{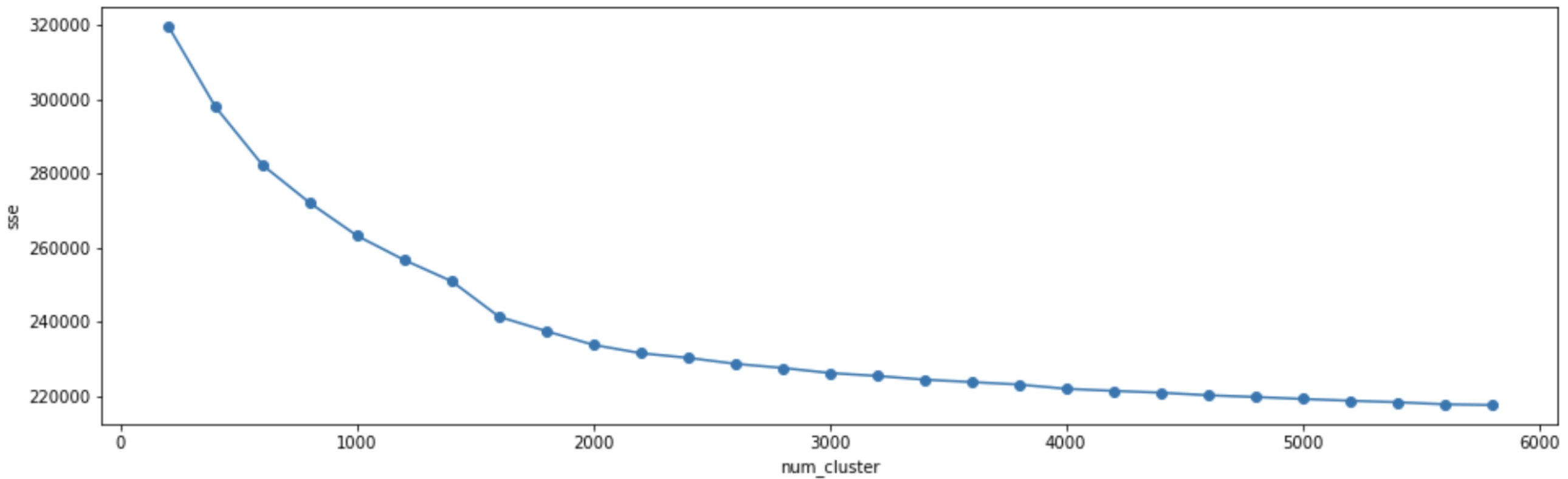}
  \caption{SSE versus the number of cluster K employed in unsupervised target domain data}
  \label{fig:sse}
\end{figure}

\section{Task3: Semi-supervised domain adaptation}

Semi-supervised domain adaptation is a new task proposed this year. We are following the FFSVC baseline system method we proposed \cite{ffsvc22}. The pseudo-label method is adopted for a large-scale set of unlabeled data from the target domain. The detail is shown in the following.

\subsection{Pseudo Label Method}

In this case, the VoxCeleb data is considered as the out-of-domain labeled data and the permitted CN-Celeb data is treated as the in-domain unlabeled data. The labeled data is used to pre-train a model for pseudo-labeling. For semi-supervised learning operation in the transfer learning phase, the speaker embedding from unlabeled data is extracted using the pre-trained model and the following clustering algorithms to generate pseudo-labels for the unlabeled data. The following describes the algorithm generating pseudo-labels:

\begin{itemize}
	\item Step 1. Cleaning audios. Since too short duration audios may not contain text information, only audios greater than 1s in length are retained.  
    \item Step 2. Extract embedding. Extract all speaker embeddings $Z \in R^{N \times d}$ from the permitted CN-Celeb dataset using the pre-trained speaker model.
	\item Step 3. Clustering. Run a clustering algorithm with the different number of clusters K to obtain centroid matrix $C\in R^{d}$ for each K.
	\item Step 4. Calculate the sum of square error (SSE) and observe the `elbow' of the SSE curve to determine the number of clusters K .
    \item Step 5. Pseudo labeling. Create the pseudo labels for the target domain unlabeled data.
    \item Step 6. Fine-tuning. Use the pseudo-labels data together with the labeled data into the speaker embedding model to fine-tune the model.
    \item Step 7. Repeat Step 2 with the fine-tuned model from Step 6 as the pre-trained model.
    
\end{itemize}

\begin{table*}[htbp]\centering \scriptsize 
    \caption{The performance of various systems in the track3.}
     \label{tab:final_system_task3}
    \begin{tabular}{lccccccc}
    \toprule
    \multirow{2}*{\textbf{ID \& Model}} & \multirow{2}*{\textbf{Iteration}} & \multicolumn{2}{c}{\textbf{VoxSRC22 val}} & \multicolumn{2}{c}{\textbf{VoxSRC22 eval}} \\
    \cmidrule(lr){3-4} \cmidrule(lr){5-6} & & \textbf{EER[\%]} & \textbf{mDCF$_{0.05}$} & \textbf{EER[\%]} & \textbf{mDCF$_{0.05}$}  \\
    \midrule 
	
    1 SimAM-ResNet34 & - &  &  & \\
    + pseudo label (ArcFace) & Round1 & 10.726 & 0.458 & - & -\\
    + pseudo label (Sub-center ArcFace) &  Round1 & 9.735 & 0.415 & 9.747 & 0.4985\\
    ++ AS-Norm & Round1 & 9.290 & 0.389 & - & - \\
    +++ QMF &  Round1 & 8.459 & 0.397 & 8.332 & 0.449\\
    + pseudo label (Sub-center ArcFace) &  Round2 & 10.010 & 0.434 & - & -\\
    \midrule
    
    2 ResNet101-ASP \\
    + pseudo label (Sub-center ArcFace) & Round1 & 8.425 & 0.385 & - & -\\
    ++ AS-Norm & Round1 & 8.275 & 0.367 & - & -\\
    +++ QMF & Round1 & 8.065 & 0.374 & - & -\\
    3 ResNet152-Stat  \\
     + pseudo label (Sub-center ArcFace) & Round1 & 8.165 & 0.375 & - & -\\
    ++ AS-Norm & Round1 & 7.875 & 0.356 & - & -\\
    +++ QMF & Round1 & 7.455 & 0.381 & - & -\\
    4 ResNet152-ASP  \\
    + pseudo label (Sub-center ArcFace) & Round1 & 8.335 & 0.369 & - & -\\
    ++ AS-Norm & Round1 & 8.020 & 0.347 & - & -\\
    +++ QMF & Round1 & 7.730 & 0.365 & - & - \\
    
    5 Res2Net101-Std \\
    + pseudo label (Sub-center ArcFace) & Round1 & 8.440 & 0.387 & - & -\\
    ++ AS-Norm & Round1 & 8.345 & 0.362 & - & -\\
    +++ QMF & Round1 & 7.810 & 0.390 & - & - \\
    
    6 SE-ResNet101-ASP \\
    + pseudo label (Sub-center ArcFace) & Round1 & 8.680 & 0.398 & - & -\\
    ++ AS-Norm & Round1 & 8.560 & 0.375 & - & -\\
    +++ QMF & Round1 & 8.345 & 0.391 & - & -\\
    
    7 ResNet221Stat \\
    + pseudo label (Sub-center ArcFace) &Round1 & 9.160 & 0.376 & - & -\\
    ++ AS-Norm & Round1 & 8.810 & 0.357 & - & -\\
    +++ QMF & Round1 & 8.090 & 0.372 & - & - \\ 
    
    8 RepVGG-ASP \\
    + pseudo label (Sub-center ArcFace) & Round1 & 8.570 & 0.382 & - & - \\
    ++ AS-Norm & Round1 & 8.445 & 0.360 & - & -\\
    +++ QMF & Round1 & 8.270 & 0.372 & - & - \\
    \midrule
    Fusion(1+2+3+4+5+6+7+8) & - & 7.000 & 0.326 & 7.153 & 0.389 \\
    
    \bottomrule
     
    \end{tabular}
    \end{table*}

\textbf{Generating pseudo label by clustering}. We adopt the K-means algorithm as the clustering algorithm to generate the pseudo labels. The learning objective of K-means is set to minimize the within-cluster sum-of-squares criterion:
\begin{equation}
   \underset{\mathbf{C}}{min}\frac{1}{N}\sum_{i=1}^{N}\underset{k}{min}\Vert \mathbf{z}_{i}-\mathbf{C}_{k}\Vert^{2}
\end{equation}
where $\mathbf{z}_{i}\in \mathbb{R}^{d}$ is the $d$-dimensional speaker embedding of the $i^{th}$ sample. The cluster with the closest centroid to $z_{i}$ in terms of the L2-norm distance is assigned as the pseudo-label for sample i. 

\textbf{Determine the number of clusters}. Inspired by the works of Cai \textit{et al.}\cite{dku_voxsrc20,dku_voxsrc21_ssl}, we determine the number of clusters by the `elbow' method. Given $\mathbf{z}_{k,a}$, the assigned $a^{th}$ embedding of the $k^{th}$ cluster. The total SSE of $N$ elements is:  

\begin{equation}
	{SSE}=\sum_{i=1}^k\sum_{a=1}^{A}\Vert\mathbf{z}_{k,a}-\mathbf{C}_{k}\Vert^{2}
\end{equation}

Fig. \ref{fig:sse} shows the curve of SSE results under different $K$s. The SSE monotonically increases as the number of clusters $K$ increases. SSE tends to flatten with some $K$ onwards and forming an `elbow' of the curve. Such `elbow' indicates that the intra-cluster has little variation and is increasingly over-fitting. From Fig. \ref{fig:sse}, the `elbow' is distributed between 1600 and 2000. The specific implementation of the baseline system has been released\footnote{https://github.com/FFSVC/FFSVC2022\_Baseline\_System}. Therefore, the experiment conducts with K=1600,1800,2000.

\subsection{Training Strategy and Implement Detail}
In this task, the training consists of two parts: pre-training and fine-tuning. The pre-training stage is the implementation of the track1. We adopt the pre-trained model used in track1. In the fine-tuning stage, different from the fine-tuning method of the FFSVC task2 baseline, only permitted CN-Celeb data is fed into the pre-trained speaker model. We also attempt to train with the mix of Vox2dev together with CN-Celeb, but the performance is decline.

In contrast to task1, we adopt more speaker embedding models in this task. Although the performance of the additional speaker model is not the best in the VoxCeleb-O, the scenarios of target domain data are complex. We hope that the different models can complement each other. Therefore, we add ResNet221-Stat, RepVGG-ASP\cite{repvgg,speakin21} and SE-ResNet101-ASP models to our candidate model list.  In addition, since clustering inevitably generates noisy labels, the ArcFace is replaced by Sub-Center ArcFace\cite{subcenter-arcface}. Since the data distribution of track3 validation is the same as the evaluation set, QMF-based score calibration is operated on the validation set. For score normalization, 20,000 utterances are randomly selected from the target domain unlabeled data as the cohort. What's more, consider the long-time utterances in validation and evaluation set, we also truncate the audio for a maximum of 20s. 

\subsection{Results}
    
Table.\ref{tab:final_system_task3} reports the system performance in track3. First, the SimAM-ResNet34 model is adopted as the baseline system. Sub-center ArcFace achieves about a relative 10\% improvement over ArcFace. Since the validation and evaluation set have the same data distribution, score calibration and normalization could significantly improve the system performance. It is worth noting that the performance decays in the second round. We also conduct the second round of fine-tuning on other models, the experimental results are consistent. Therefore, all semi-supervised speaker embedding models only operate one round of clustering and fine-tuning. The final submitted results achieve the 7.153\% EER in the evaluation set. 

\section{Conclusions}
This paper describes the our system in Track 1 and 3. For track 3, the pseudo-label method is introduced in this challenge to make semi-supervised domain adaptation. To mitigate the noisy label generated by clustering, the Sub-Center ArcFace is adopted to replace the ArcFace. After the AS-Norm and score calibration operation, the final fusion system achieves the 7.153\% EER in the evaluation set.

\bibliographystyle{IEEEtran}

\bibliography{mybib}

\end{document}